\documentclass[runningheads]{svmult}

\usepackage{graphicx}  
\usepackage{subeqnar}  
\usepackage{physprbb}  

\begin{document}
\title*{Close Companions to Nearby Young Stars\protect\newline from Adaptive Optics Imaging on VLT and Keck}
\toctitle{Close Companions to Nearby Young Stars
\protect\newline from AO on VLT and Keck}
%
%
\titlerunning{Close Companions to Nearby Young Stars with AO}
%
\author{Karl E. Haisch Jr.\inst{1}
\and Ray Jayawardhana\inst{1}
\and Alexis Brandeker\inst{2}
\and Diego Mardones\inst{3}}
%
\authorrunning{Haisch Jr. et al.}
%
%
\institute{University of Michigan, Ann Arbor, Michigan 48109-1090, U.S.A.
\and Stockholm Observatory,
     AlbaNova University Center,
     SE-106 91 Stockholm, Sweden
\and Departamento de Astronomia, Universidad de Chile, 
	Casilla 36-D, Santiago, Chile}

\maketitle              

\begin{abstract}
We report the results of VLT and Keck adaptive optics surveys of known members of the $\eta$ Chamaeleontis, MBM 12, and TW Hydrae (TWA) associations to search for close companions. The multiplicity statistics of $\eta$ Cha, MBM 12, and TWA are quite high compared with other clusters and associations, although our errors are large due to small number statistics. We have resolved S18 in MBM 12
and RECX 9 in $\eta$ Cha into triples for the first time. The tight binary TWA 5Aab in the TWA offers the prospect of measuring the dynamical masses of both components as well as an independent distance to the system within a few years. The AO detection of the close companion to the nearby young star $\chi$$^{1}$ Orionis, previously inferred from radial velocity and astrometric observations, has already made it possible to derive the dynamical masses of that system without any astrophysical assumption.
\end{abstract}

\section{Introduction}
Nearby, young stellar associations offer unique advantages for detailed studies of star and planet formation. With diffraction limited adaptive optics (AO) near-infrared observations on 8 - 10 m class telescopes, their proximity facilitates sensitive studies of individual star systems down to 40 - 50 milliarcsecond scales (corresponding to physical scales of $\sim$ 2 - 3 AU at 50 pc, or 6 - 9 AU at 150 pc). Resolutions of a few AU orbital radius imply orbital periods of only a few years, closing the gap between spectroscopic and visual binaries and thereby offering the prospect of a complete census of multiplicity among young pre-main-sequence (PMS) stars down to substellar masses. Resolving companions spatially is important since spectroscopic observations by themselves only yield the relative masses of binaries. By achieving resolutions down to a physical scale of a few AU, it is not only possible to determine the dynamical masses of the components, but also to derive an independent distance to the system. Thus, spatially resolved PMS binaries are essential tools for testing models of PMS stellar structure and evolution.

Searching for companions of PMS stars is also important since this allows us to address the ambiguity when placing them on evolutionary luminosity-color diagrams. Unresolved PMS binaries show luminosity (and possibly color) differences compared to the individual components, thereby biasing any conclusions drawn from evolutionary diagrams (see \cite{math94} for an extensive review).

The nearby young associations $\eta$ Chamaeleontis ($d$ $\sim$ 97pc; \cite{mam99}), MBM 12 ($d$ $\sim$ 275pc; \cite{luh01}), and TWA ($d$ $\sim$ 55 pc; \cite{kast97}, \cite{per97}) are prime targets for detailed studies of close companions. Here, we report the results of an AO survey of known members of these associations.

\section{Observations and Detection Limits}

Near-infrared AO observations of nine members of the $\eta$ Cha association were acquired with the 8.2m VLT UT4 telescope on Cerro Paranal, Chile using NAOS-CONICA (NACO). We surveyed an additional 21 young stars (9 in MBM 12 and 12 in TWA) with the 10m Keck II telescope on Mauna Kea, Hawaii using AO \cite{win00} and the near-infrared camera KCam. The pixel scales of NACO and KCam are 13.26 mas pixel$^{-1}$ and 17.47 mas pixel$^{-1}$ respectively. The science target itself was used as a wavefront sensor for the AO system. Our KCam observations were acquired in three filters, $J$(1.26 $\mu$m), $H$(1.648 $\mu$m), and $K^\prime$(2.127$\mu$m), while only the $H$-band was used for the NACO observations. The Strehl ratio in all cases was 0.2 - 0.25 in $H$-band.

\begin{figure}[h]
\begin{center}
\includegraphics[width=.5\textwidth]{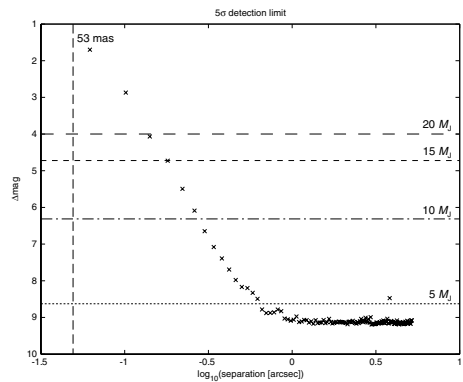}
\end{center}
\caption[]{Approximate 5$\sigma$ detection contrast sensitivity as a function of separation for RECX 9 in $\eta$ Cha. Horizontal lines correspond to brown dwarfs having masses of 5, 10, 15, and 20 Jupiter masses as determined from models of 10 Myr brown dwarfs from \cite{bar03}}
\label{fig1}
\end{figure}

In order to estimate our sensitivity to finding close, faint companions, the speckle noise was measured from observed point-spread functions of single stars. Our approximate 5$\sigma$ detection limit as a function of separation for RECX 9 in $\eta$ Cha from VLT/NACO is shown in Figure~\ref{fig1}. The locations of brown dwarfs having masses of 5, 10, 15, and 20 Jupiter masses as determined from models of 10 Myr brown dwarfs from \cite{bar03} are indicated in the figure.

\section{Results}

\subsection{Companions}

\begin{figure}
\begin{center}
\leavevmode
\includegraphics[width=.5\textwidth]{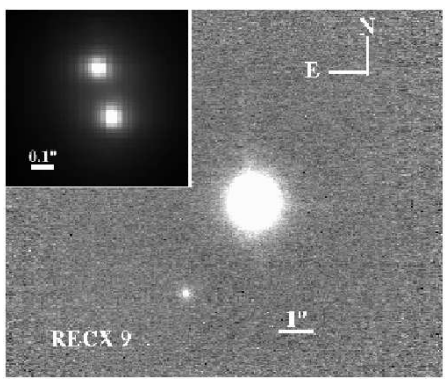}
\end{center}
\caption{$H$-band image of $\eta$ Cha member RECX 9. The image scale is 13.26 mas pixel$^{-1}$. North is up, and east is to the left. RECX 9 is a new triple system previously known as a tight $\sim$ 0.25 arcsec binary \cite{kpg02} (see inset)}
\label{fig2}
\end{figure}

Five of the nine $\eta$ Cha members surveyed were found to be candidate multiple objects, including four new binaries (RECX 3, RECX 4, RECX 10, and ECHA J0843.3-7905 with separations of 2.04, 7.35, 9.89, and 6.35 arcseconds respectively), and one new triple (RECX 9 [see Figure~\ref{fig2}]; which was previously known as a tight $\sim$ 0.25 arcsec binary \cite{kpg02}). RECX 12 appears elongated and is suspected to be a binary, however it is unresolved in our images. The suspected binary nature of RECX 12 has also been noted by \cite{kpg02}.

Of the nine members observed in the MBM 12 association, five were found to be binaries, all previously reported by \cite{cha02}, and one source, S18, is a new triple (with a 63 mas, $\sim$ 17 AU, binary). In TWA, no new companions were found apart from TWA 5Aab (54 mas, $\sim$ 3 AU, separation) which was independently discovered by \cite{mac01}. Images of S18 and TWA 5Aab can be found in \cite{bra03}. 

The separation of TWA 5Aab, in conjunction with Kepler's law and assuming a circular orbit, implies an orbital period of $P$ = 5.2($M_B$$\alpha$$^{3}$)$^{-1/2}$ yr, where $\alpha$ is the projection factor [observed (projected) separation/real separation], and $M_B$ is the binary system mass in solar units. This corresponds to an average position angle motion of 69($M_B$$\alpha$$^{3}$)$^{1/2}$ deg yr$^{-1}$. The radial relative velocity amplitude is 17.3($M_A$$\alpha$)$^{1/2}$ sin $i$ km s$^{-1}$. Thus, within a few years, subsequent studies have the potential to yield accurate dynamical mass estimates for the individual components of this system.

\subsection{Multiplicity Statistics}

The binary fraction (BF) and companion star fraction (CSF) are defined as:
\begin{subeqnarray}
BF & = & \frac {\it B + T + Q}{\it S + B + T + Q} \; , \label{f1a}\\
CSF & = & \frac {\it B + 2T + 3Q}{\it S + B + T + Q} \;  .\label{f1b}
\end{subeqnarray}

\noindent where $S$ is the number of single stars, $B$ is the number of binary systems, $T$, the number of triple systems, and $Q$, the number of quadruple systems. In Table~\ref{Tab1}, we summarize, for each region surveyed, the BFs, and CSFs for physical separations ranging from $\sim$ 15 - 600 AU. The quoted uncertainties in the CSFs represent the statistical $\sqrt{N}$ errors. The multiplicity statistics for $\eta$ Cha, MBM 12, and TWA are rather high compared with those reported in other young clusters and associations (CSF $<$ 0.60; \cite{duc99}), however the significance of this result is low due to our small number statistics.

\begin{table}
\caption{Binary and Companion Star Fractions}
\begin{center}
\renewcommand{\arraystretch}{1.4}
\setlength\tabcolsep{15pt}
\begin{tabular}{@{}lcc}
\hline\noalign{\smallskip}
Association & BF & CSF \\
\noalign{\smallskip}
\hline
\noalign{\smallskip}
MBM 12 & 0.64 $\pm$ 0.16 & 0.91 $\pm$ 0.30 \\
TWA & 0.58 $\pm$ 0.12 & 0.84 $\pm$ 0.22 \\
$\eta$ Cha & 0.56 $\pm$ 0.25 & 0.67 $\pm$ 0.27 \\
\noalign{\smallskip}
\hline
\noalign{\smallskip}
\end{tabular}
\end{center}
\label{Tab1}
\end{table}

\section{$\chi$$^{1}$ Orionis: Direct Determination of Component Masses}

The G0V-star $\chi$$^{1}$ Ori is a known single-lined spectroscopic and astrometric binary, the orbital parameters of which were first derived by \cite{lw78}. Using their astrometric data and radial velocity data from \cite{mb92} and \cite{mb98}, \cite{hg02} derived a period $P$ = 5156 $\pm$ 2.5 days, and a mass ratio $q$ = $M_B$/$M_A$ = 0.15 $\pm$ 0.005.

\begin{figure}[h]
\begin{center}
\includegraphics[width=.48\textwidth]{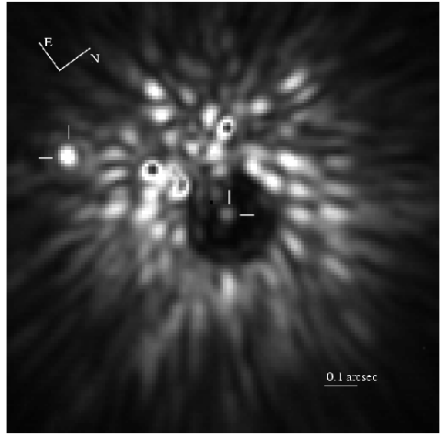}
\end{center}
\caption[]{$H$-band image of $\chi$$^{1}$ Ori and its companion. North and East are as labelled in the image \cite{kon02}}
\label{fig3}
\end{figure}

Recently, \cite{kon02} used the AO system on the Keck II telescope, the NIRC2 camera, and a 300 mas diameter coronograph to obtain an $H$-band image of $\chi$$^{1}$ Ori (Figure~\ref{fig3}). The coronograph is semi-transparent, with a throughput slightly below 0.5\%, so the position of the star behind it can be measured precisely. The companion is clearly visible in the 0.18 second image. Knowledge of the Hipparchos parallax for $\chi$$^{1}$ Ori of 115.43 $\pm$ 1.08 mas allowed the absolute $H$-band magnitude of the secondary to be calculated. Combined with the orbital elements published by \cite{hg02}, \cite{kon02} calculated the dynamical masses of both components of the $\chi$$^{1}$ Ori binary directly for the first time without any astrophysical assumption using Kepler's laws. The final masses are M$_A$ = 1.01 $\pm$ 0.13 M$_\odot$ and M$_B$ = 0.15 $\pm$ 0.02 M$_\odot$.

\section{Conclusions}

We report the results of an AO survey of the known members of the MBM 12, TW Hydrae, and $\eta$ Chamaeleontis associations to search for close companions. The main scientific conclusions can be summarized as follows:

\begin{itemize}
\item
Near-infrared AO observations on 8 - 10 meter class telescopes are closing the sensitivity gap to the spectroscopic surveys for binaries and are allowing direct component dynamical mass estimates\\
\item
The multiplicity statistics of $\eta$ Cha, MBM 12, and TWA are quite high compared with other clusters and associations, although the significance of this result is low due to our small number statistics\\
\item
Within a few years, follow-up astrometric and spectroscopic observations of TWA 5A have the potential to provide both the dynamical masses of the individual components of this tight 54 mas binary, as well as an independent distance to the system\\
\item
AO observations have allowed the individual component dynamical masses of $\chi$$^{1}$ Ori to be calculated for the first time without any astrophysical assumptions
\end{itemize}

%

\end{document}